\documentclass[twocolumn,twoside,slac_two]{revtex4}
\usepackage{graphicx}
\usepackage{fancyhdr}
\begin{document}

\title{First Observation of Accelerator Muon Antineutrinos in MINOS}

\author{Istvan Danko (for the MINOS Collaboration)}
\affiliation{Department of Physics and Astronomy, University of Pittsburgh, Pittsburgh, PA 15260, USA}

\begin{abstract}
We report the first direct observation of muon antineutrinos in the MINOS Far Detector in the current muon-neutrino dominated beam. The magnetic field of the detector is utilized to separate muon neutrinos and antineutrinos event-by-event by identifying the charge sign of the muon created in charged-current interactions. We present preliminary results on the ${\bar \nu}_{\mu}$ oscillation parameters as well as limit on the fraction of neutrinos that disappear and reappear as antineutrinos. We also discuss the prospect of the measurement when the polarity of the magnetic focusing horns will be reversed to create a dedicated muon antineutrino beam.
\end{abstract}

\maketitle

\thispagestyle{fancy}

\section{Introduction}

MINOS is a long-baseline neutrino experiment in the NuMI (Neutrinos at the Main Injector) beam line at Fermilab that has confirmed the disappearance of muon neutrinos and measured the atmospheric oscillation parameters ($\sin^2(2\theta)$ and $|\Delta m^2|$) with high precision \cite{MINOS_CC}. In this paper, we describe a new study that utilizes the 7\% muon antineutrino component of the beam to measure their oscillation parameters directly and test exotic models such as CPT violation in the neutrino sector \cite{CPT} and transition of $\nu_{\mu}$ to ${\bar \nu}_{\mu}$ \cite{nu_transition}.

In a quasi-two-neutrino mixing framework, ${\bar \nu}_{\mu}$ oscillation implies a survival probability of
\begin{equation}
P( {\bar \nu}_{\mu} \to {\bar \nu}_{\mu}) = 1 - \sin^2 (2{\bar \theta}) \sin^2 \left(\frac{1.27 \Delta {\bar m}^2 L}{E} \right),
\label{eq:oscillation}
\end{equation}
where $E$ is the neutrino energy in GeV, $L$ is the distance traveled by the neutrino in km, and the antineutrino mixing angle (${\bar \theta}$) and mass separation ($\Delta {\bar m}^2$) are assumed to be independent of the corresponding neutrino parameters\footnote{In this framework, $\nu_{\mu}$ oscillation is dominated by the mixing between the $\nu_2$ and $\nu_3$ mass eigenstates, thus $\Delta m^2 \approx \Delta m^2_{23} = m^2_3 - m^2_2$ and $\theta \approx \theta_{23}$. Similarly for the antineutrinos.} ($\theta$ and $\Delta m^2$). 
In addition to the oscillation scenario, we also consider the possibility that a fraction, $\alpha$, of the muon neutrinos that have been observed to disappear along their long flight \cite{MINOS_CC} will reappear as antineutrinos. The probability of this transition can be parametrized empirically as
\begin{equation}
P( \nu_{\mu} \to {\bar \nu}_{\mu}) = \alpha \sin^2 (2\theta) \sin^2 \left(\frac{1.27 \Delta m^2 L}{E} \right).
\label{eq:transition}
\end{equation}

\section{Neutrino Beam and Detectors}

MINOS uses the high intensity neutrino beam \cite{NuMI} created by 120 GeV protons from Fermilab's Main Injector impinged on a graphite target. Secondary particles, mainly $\pi$ and $K$ mesons, from the target are sign-selected and focused by two toroidal magnetic horns toward a 675 m long evacuated pipe where they can decay in flight to produce neutrinos. In the current beam configuration the horns are focusing (defocusing) secondary particles with positive (negative) charge, which enhances the fraction of muon neutrinos at the expense of antineutrinos. Most of the ${\bar \nu}_{\mu}$ arises from $\pi^-$ (and $K^-$) parents produced \emph{upstream} in the target that travel down the center of the horns where they are not deflected by the magnetic field. In addition, a significant portion of the ${\bar \nu}_{\mu}$ originate from parents produced \emph{downstream} from the target in secondary interactions in the decay pipe wall and the surrounding material.

The MINOS experiment \cite{MINOS} uses two functionally identical scintillator tracking/sampling-calorimeter detectors located 1 km (Near Detector) and 735 km (Far Detector) from the target. This two-detector setup significantly reduces the systematic uncertainties due to neutrino flux, cross section, and detection efficiency. The detector design is optimized to observe charged-current (CC) muon (anti)neutrino interactions which produce a prominent muon track penetrating several layers of alternating iron and scintillator planes. Both detectors have a toroidal magnetic field with an average field strength of about 1.3 T (near) and 1.4 T (far) in the steel. The momentum of the muons is measured from their range in the detector or alternatively from their curvature in the magnetic field. In addition, positive (negative) muons are focused (defocused) by the field allowing to discriminate between $\nu_{\mu}$-CC and ${\bar \nu}_{\mu}$-CC interactions event-by-event. 

\begin{figure}[h]
\centering
\includegraphics[width=80mm]{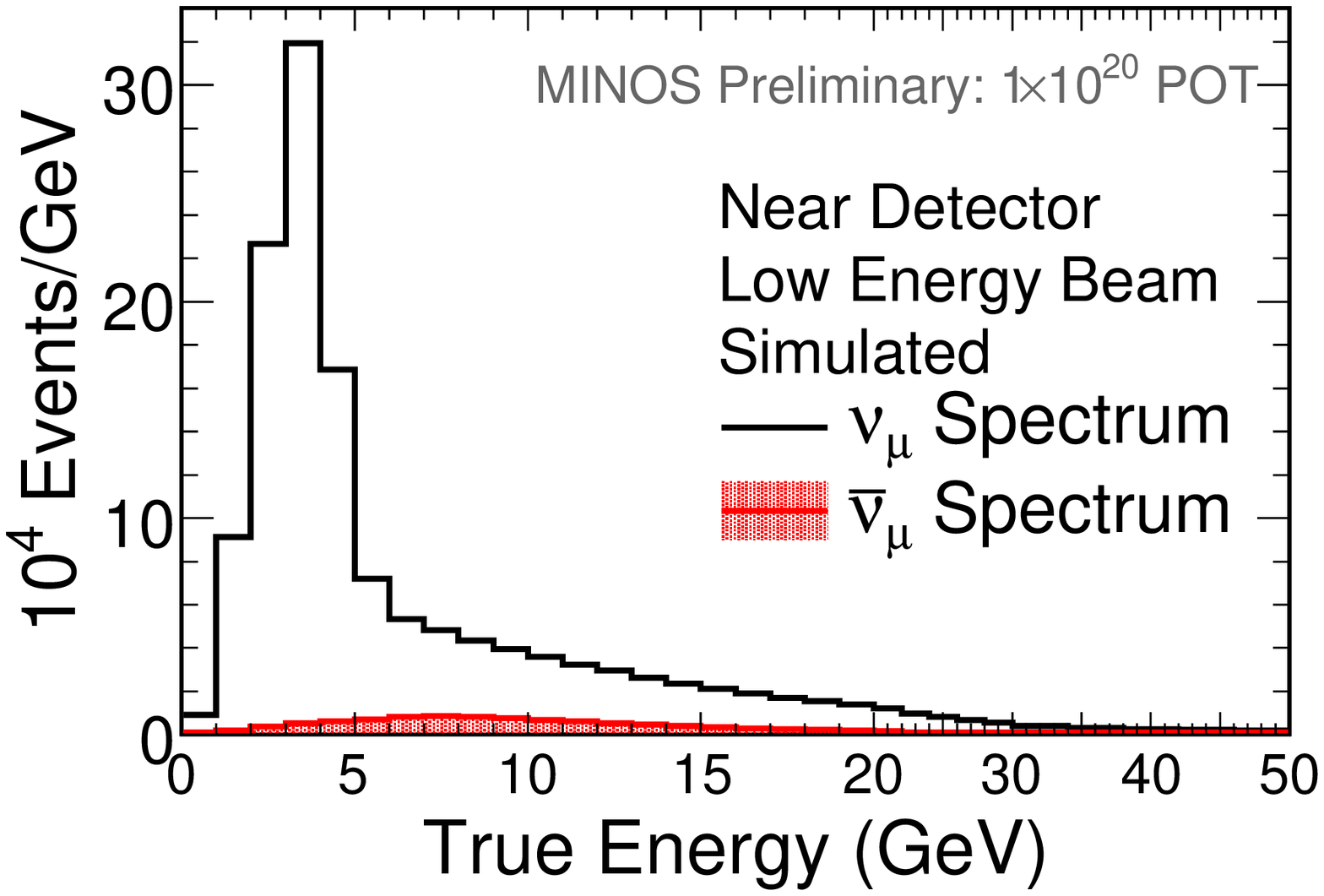}
\includegraphics[width=80mm]{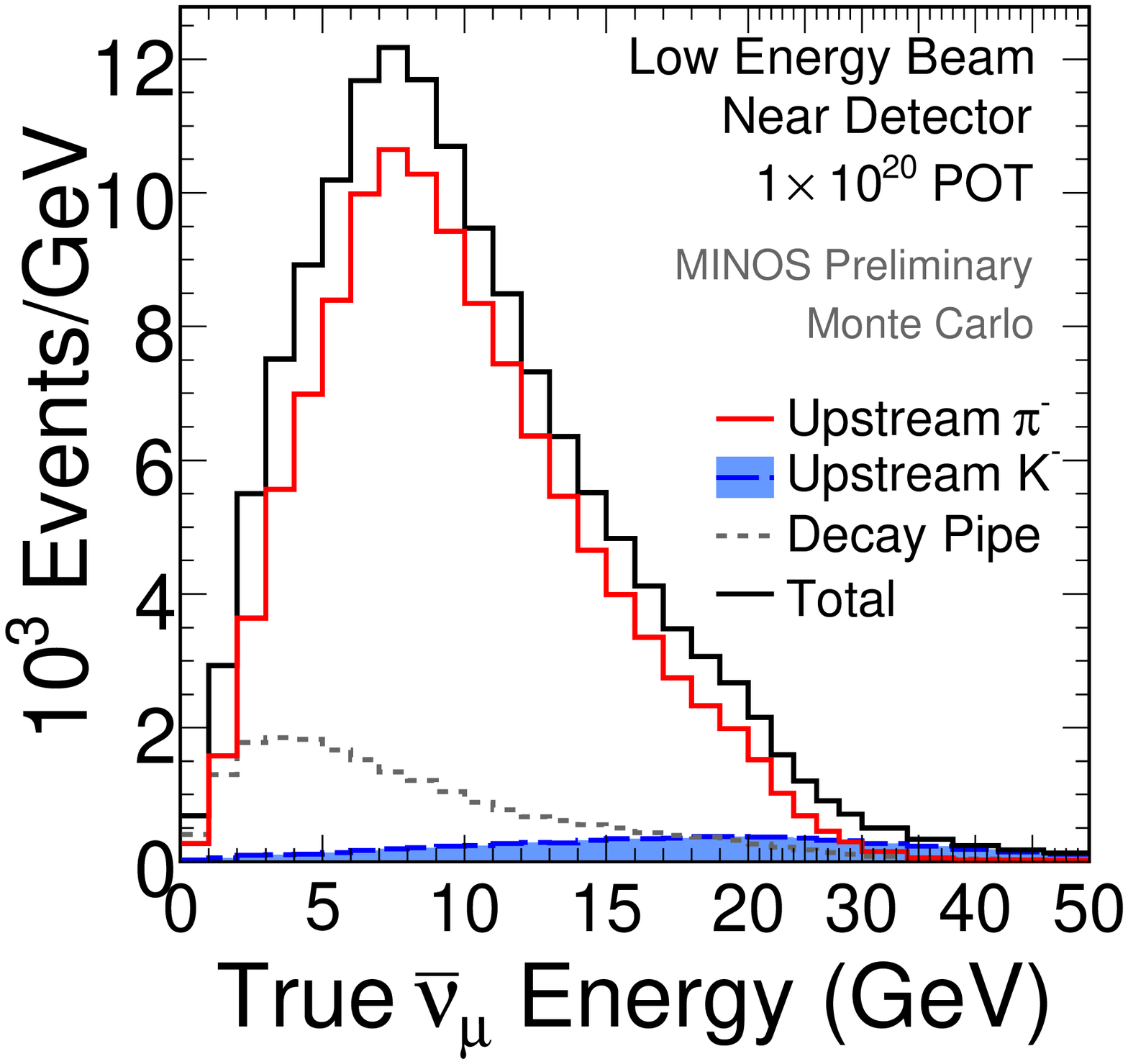}
\caption{Top: the relative size of the ${\bar \nu}_{\mu}$-CC energy spectrum (red) compared to the $\nu_{\mu}$-CC spectrum (black) in the Near Detector. Bottom: the contribution of the different beam components to the Near Detector ${\bar \nu}_{\mu}$-CC spectrum.}
\label{fig:Flux}
\end{figure}

Fig.~\ref{fig:Flux} shows the predicted number of $\nu_{\mu}$ and ${\bar \nu}_{\mu}$ CC interactions in the MINOS Near Detector and the contribution of different beam components to the  ${\bar \nu}_{\mu}$-CC spectrum. Due to the focusing/defocusing effects of the horns on the parent mesons, the shape of the $\nu_{\mu}$ and ${\bar \nu}_{\mu}$ spectra are significantly different. The higher peak energy of the ${\bar \nu}_{\mu}$ spectrum around 8 GeV makes the ${\bar \nu}_{\mu}$ measurement more sensitive to higher $\Delta {\bar m}^2$ values than the $\nu_{\mu}$.

\section{Data Analysis}

The data used in this analysis corresponds to $3.2 \times 10^{20}$ protons on target (PoT) collected in the low-energy beam configuration\footnote{The peak energy of the neutrino beam can be tuned by adjusting the relative position of the target and magnetic horns as well as by changing the horn currents.} between 2005-2007.
We measure the rate of inclusive ${\bar \nu}_{\mu}$-CC interactions as a function of reconstructed antineutrino energy in the Near Detector and extrapolate the spectrum to the Far Detector. ${\bar \nu}_{\mu}$ oscillation would produce an energy-dependent deficit while $\nu_\mu \to {\bar \nu}_{\mu}$ transition would cause an excess in the Far Detector compared to the expectation.

\subsection{Event Selection}

We select events with at least one reconstructed track, the longest one identified as the muon candidate. The neutrino interaction point (vertex) is required to be inside the fiducial volume in order to suppress cosmic-ray and rock muons originating from outside of the detector as well as to contain the hadronic shower in the detector. Cosmic-ray background is suppressed further in the far detector by requiring the angle of the muon momentum to be within 53$^{\rm o}$ ($\cos\vartheta > 0.6$) of the neutrino direction and the event time to fall within 14 $\mu$s window around the time of the beam spill.

\begin{figure}[h]
\centering
\includegraphics[width=80mm]{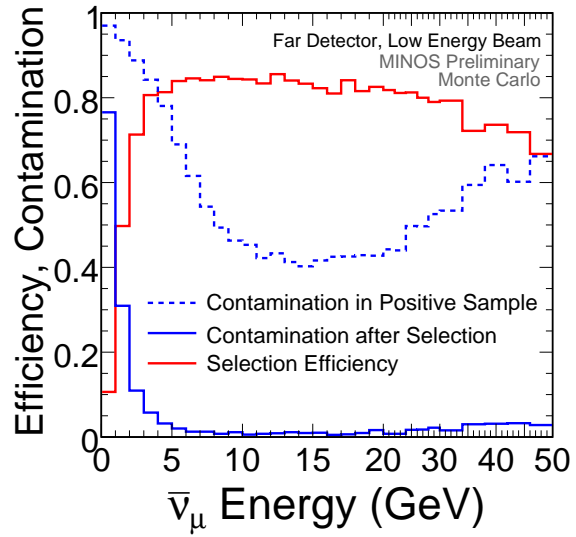}
\caption{The efficiency (solid red) and contamination (solid blue) of the final ${\bar \nu}_{\mu}$-event selection according to Monte-Carlo simulation of the far detector. The contamination in the sample with a simple selection on the charge of the muon candidate from the track fit is also shown (dashed blue).}
\label{fig:FDeffcontam}
\end{figure}

The muon candidate is required to have a positive charge determined from the direction of bending in the magnetic field. As Fig.~\ref{fig:FDeffcontam} demonstrates, this simple charge-sign selection produces a sample which is highly contaminated. The background is composed of both neutral-current (NC) events in which one or more tracks are found by the reconstruction code and high inelasticity $\nu_\mu$-CC events in which a low energy $\mu^-$ track is obscured by the hadronic shower. In addition, as the momentum of the muon increases it bends less making the charge determination harder and leading to an increasing $\nu_{\mu}$ contamination at high energy.

\begin{figure}[h]
\centering
\includegraphics[width=80mm]{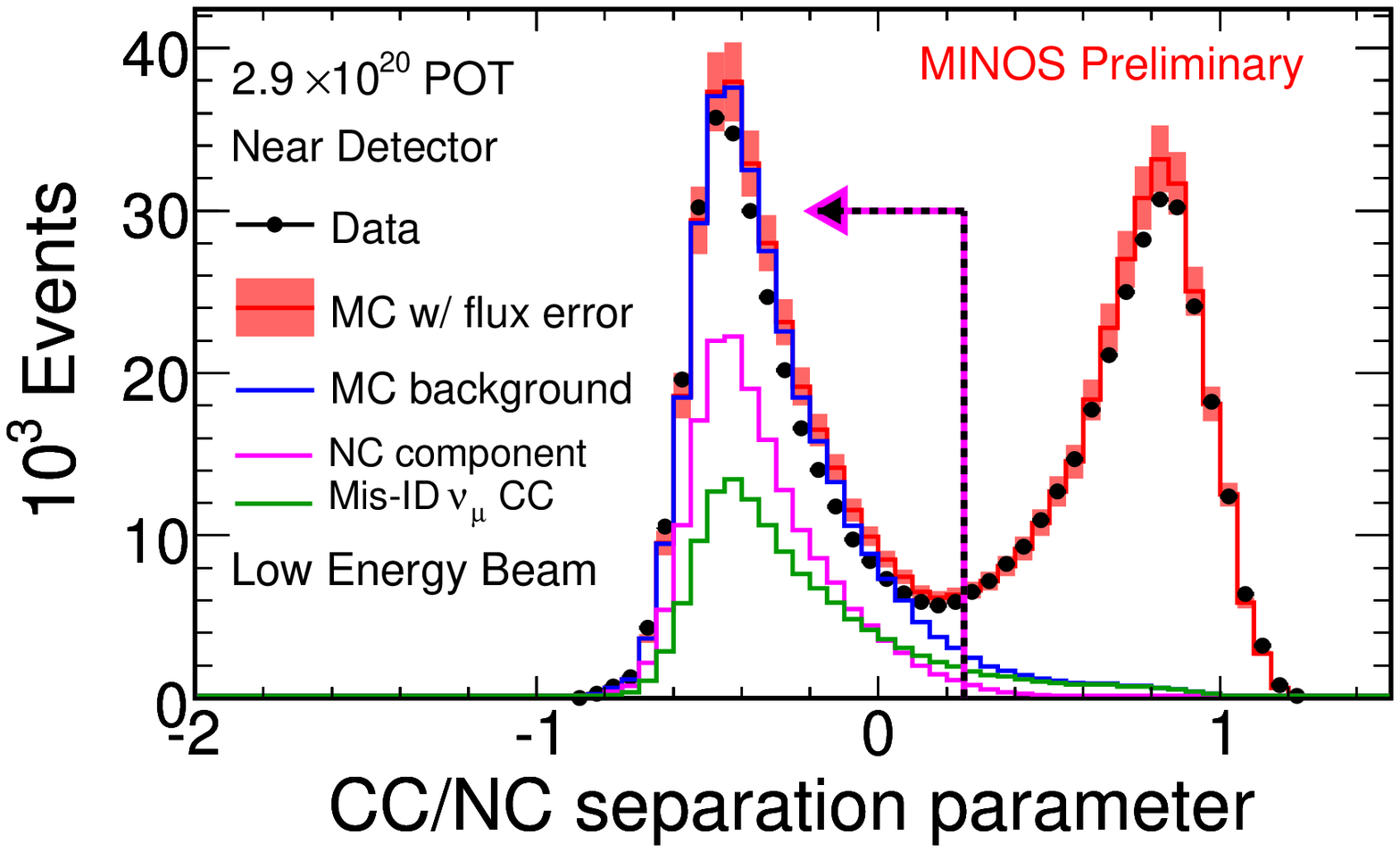}
\includegraphics[width=80mm]{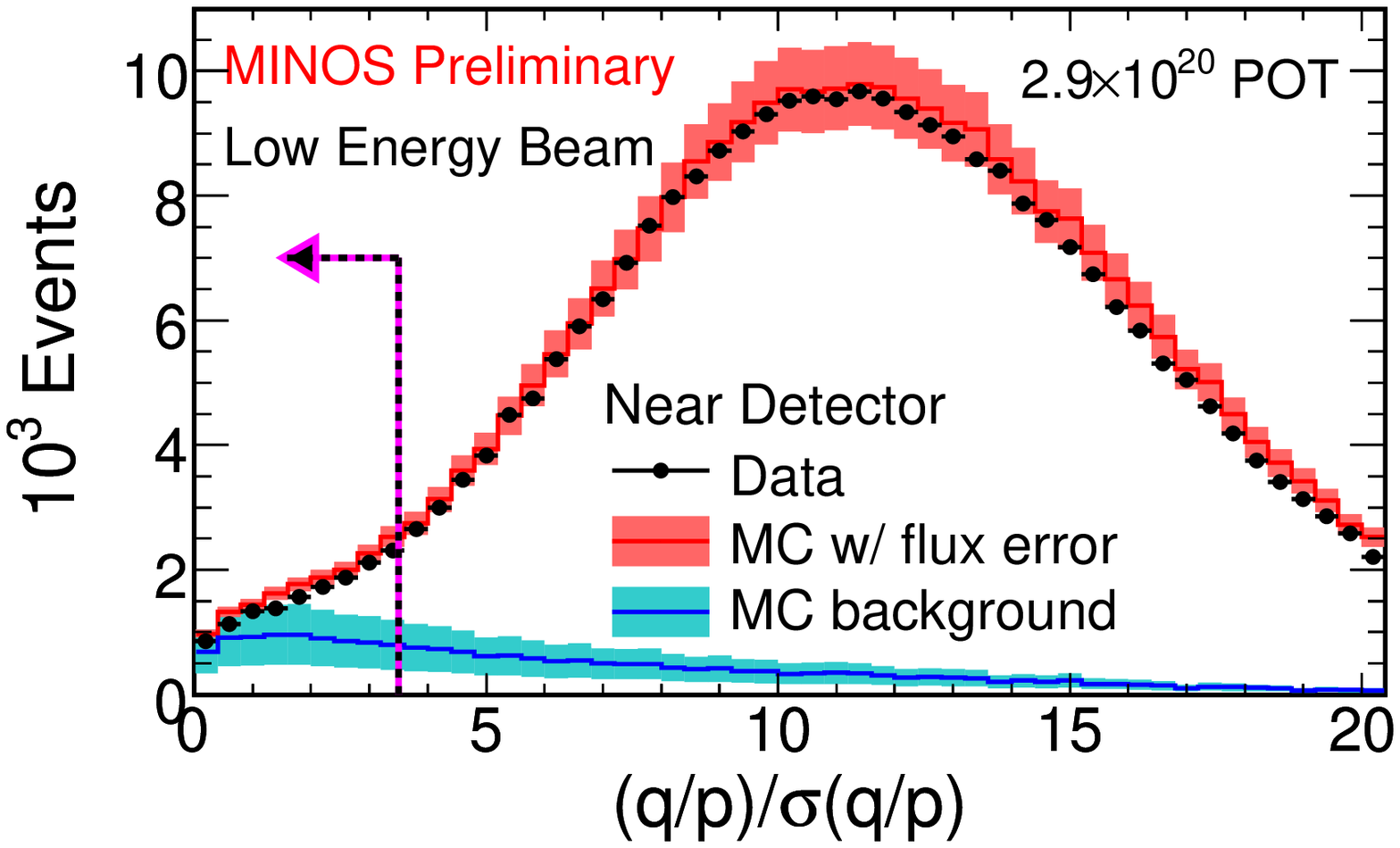}
\includegraphics[width=80mm]{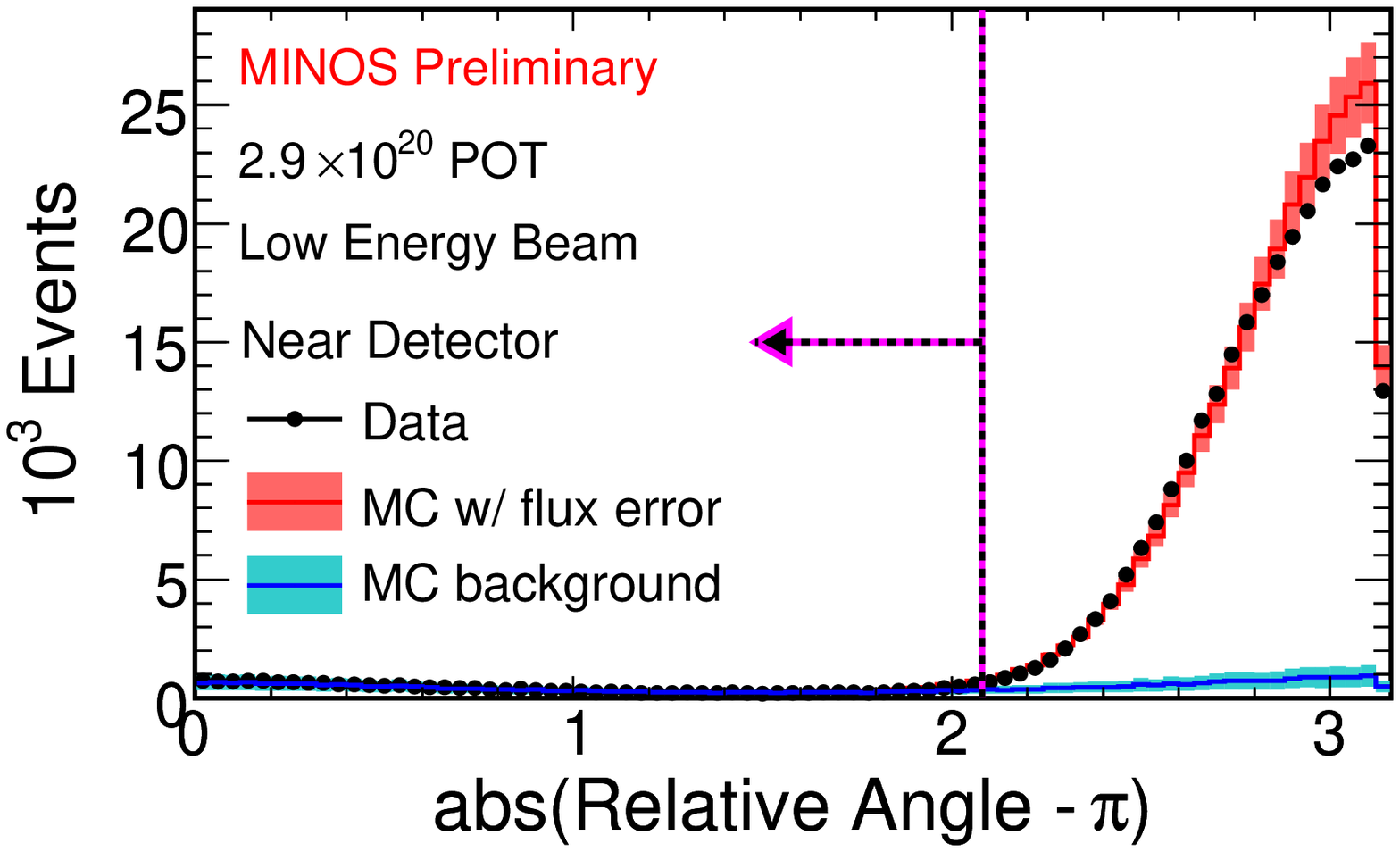}
\caption{The three variables used for ${\bar \nu}_{\mu}$ selection: the NC and CC selector (top), significance of the fit curvature (middle), and the relative angle described in the text (bottom). Each variable is shown after selection on the others are applied. The vertical lines with arrow denote the optimized selection values.} 
\label{fig:Selection}
\end{figure}

Three additional selection variables are used in order to suppress both the NC and mis-identified $\nu_{\mu}$-CC background. The first variable is a likelihood-based selector that combines three probability distribution functions related to the event topology \cite{MINOS_PRD}: event length, fraction of total event signal produced by the muon candidate, and average signal per plane induced by the muon candidate. As Fig.~\ref{fig:Selection} demonstrates, this variable is very effective to discriminate CC events with a muon track from NC events with a diffuse hadronic shower as well as the those $\nu_{\mu}$-CC events that pass the other requirements. The other two variables are designed to improve the charge-sign determination of the muon candidate. The first variable is the significance of the measured curvature from the fit, and the other is the angle defined in the plain transverse to the beam between the direction of the last muon hit with respect to the projected hit without magnetic field and the line going through the center of the magnet coil and the interaction vertex~\cite{RelAng}.

The selection was optimized for maximum efficiency $\times$ purity below 10 GeV neutrino energy in order to increase the sensitivity for CPT-conserving oscillation. Fig.~\ref{fig:FDeffcontam} shows the relative efficiency of the full selection and the remaining contamination as a function of ${\bar \nu}_{\mu}$ energy in the Far Detector. According to the MC simulation, the overall selection efficiency is 83\% with a purity of 97\% in the Far Detector and $\nu_{\mu}$-CC events are suppressed by a factor of $1.3 \times 10^{-3}$.

\subsection{Near-to-Far Extrapolation}

Even in the absence of oscillation or transition, the shape of the (anti)neutrino spectra at the Near and Far Detector are not identical due the different solid-angle coverage of the two detectors and the fact that the Near Detector sees an extended source of neutrinos while the Far Detector essentially sees a point source. The beam-line geometry and the meson decay kinematics are encapsulated in a beam-transfer matrix that relates the Near Detector energy spectrum to the Far Detector spectrum \cite{MINOS_PRD}. Using this matrix the measured Near Detector spectrum can be extrapolated to get a Far Detector prediction, which is less sensitive to uncertainties in the neutrino flux calculation. Separate transfer matrices are used for neutrinos and antineutrinos to predict the signal and background simultaneously \cite{UKMM}. Simulation of the detector response is used to correct for energy resolution as well as differences in selection efficiency and contamination between the two detectors. 

\begin{figure}[h]
\centering
\includegraphics[width=80mm]{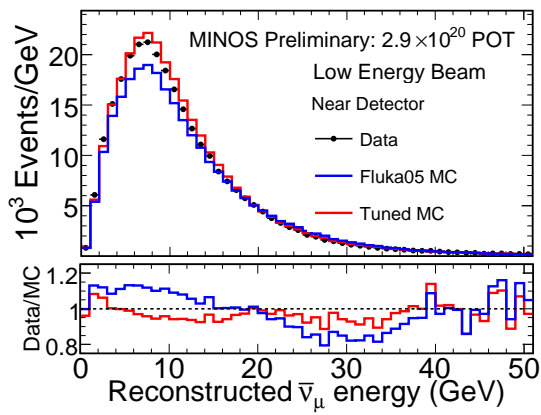}
\caption{The observed ${\bar \nu}_{\mu}$-CC spectrum in the Near Detector and the simulated spectrum before (blue) and after (red) tuning the meson production from the target (top); and the ratio of the predictions to the data (bottom).} 
\label{fig:NuMuBar_tuning}
\end{figure}

In order to improve the extrapolation further, the Near Detector data are used to constrain the hadron production off the target, which has the largest contribution to the uncertainty in the neutrino-flux calculation.
The production of $\pi^+$ and $K^+$ is parametrized as a function of their initial transverse ($p_T$) and longitudinal ($p_z$) momenta and constrained by fits to the Near Detector $\nu_{\mu}$-CC spectra measured at different beam configurations \cite{MINOS_PRD}. 
In contrast, the ${\bar \nu}_{\mu}$-CC spectrum is less sensitive to the beam configuration since the majority of ${\bar \nu}_{\mu}$ arises from low-$p_T$ parents (mainly $\pi^-$). Therefore, the $p_T$ shape of $\pi^-$ parents is constrained by the NA49 measurement of the relative $\pi^+/\pi^-$ production \cite{NA49}, while the parent $p_z$ shape and the absolute normalization are obtained from the fit to the Near Detector ${\bar \nu}_{\mu}$-CC spectrum. The result of the tuning on the antineutrino spectrum is demonstrated in Fig.~\ref{fig:NuMuBar_tuning}.

\section{Results}

\begin{figure}[h]
\centering
\includegraphics[width=80mm]{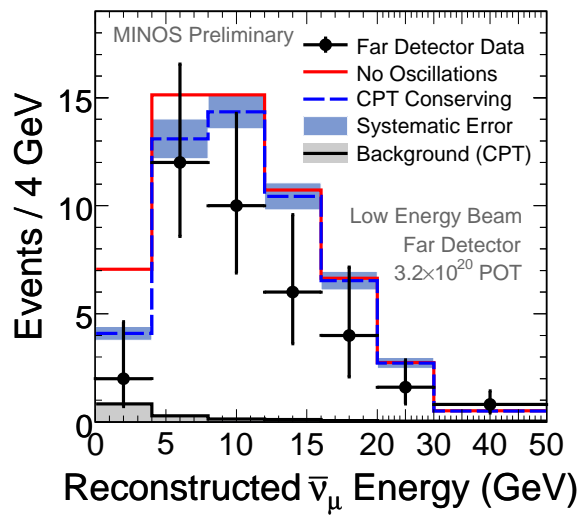}
\caption{The observed ${\bar \nu}_{\mu}$-CC spectrum (dots with error bard) and predictions in the MINOS Far Detector. The red histogram represents the prediction with no oscillation while the dashed blue with CPT-conserving oscillation. The shaded histogram is the expected background in the latter prediction.} 
\label{fig:NuMuBar_Result}
\end{figure}

Fig.~\ref{fig:NuMuBar_Result} shows the energy distribution of observed and expected ${\bar \nu}_{\mu}$-CC events in the MINOS Far Detector. We observe a total of 42 candidate events while expecting $64.6 \pm 8.0{\rm (stat.)} \pm 3.9{\rm (syst.)}$ events with no oscillation and $58.3 \pm 7.6{\rm (stat.)} \pm 3.6{\rm (syst.)}$ events with CPT-conserving oscillations when the antineutrino oscillation parameters are assumed to be equal to the best-fit MINOS neutrino oscillation parameters ($\sin^2(2{\bar \theta}) = \sin^2(2\theta) = 1$ and $\Delta {\bar m}^2 = \Delta m^2 = 2.43 \times 10^{-3}$ eV$^2$). The shaded blue band around the CPT-conserving prediction represents the systematic uncertainty, which is dominated by the relative normalization uncertainty between the two detectors, the muon momentum measurement from curvature, and the uncertainty in the contribution of downstream (decay pipe) production to the ${\bar \nu}_{\mu}$ flux. The total systematic uncertainty is less than $10$\% over the whole energy region.

Extensive consistency checks were performed to make sure the observed deficit between the data and the prediction is not due to detector, reconstruction or selection inefficiency. The whole analysis procedure was performed with an alternative event selection using independent variables and two different near-to-far extrapolation methods which resulted in a similar deficit and result. The track finding efficiency, in particular for exiting tracks, was checked using stopped and through going cosmic muons. The muon charge-sign assignment was checked using an alternative track fitter. In addition, all far detector events with a $\mu^+$ candidate, events with a $\mu^-$ candidate ending close to the detector edge, and events with no reconstructed track were visually scanned for any sign of reconstruction pathology.

The effect of neutrino oscillation or transition can be applied to the Far Detector prediction using the parametrization of Eq.~\ref{eq:oscillation} or \ref{eq:transition}, respectively, and then fitted to the data to maximize the log-likelihood ratio with respect to the oscillation parameters or $\alpha$. Confidence limits on the fit parameters are extracted using the Feldman-Cousins method \cite{FC} with systematic uncertainties incorporated.

\begin{figure}[h]
\centering
\includegraphics[width=80mm]{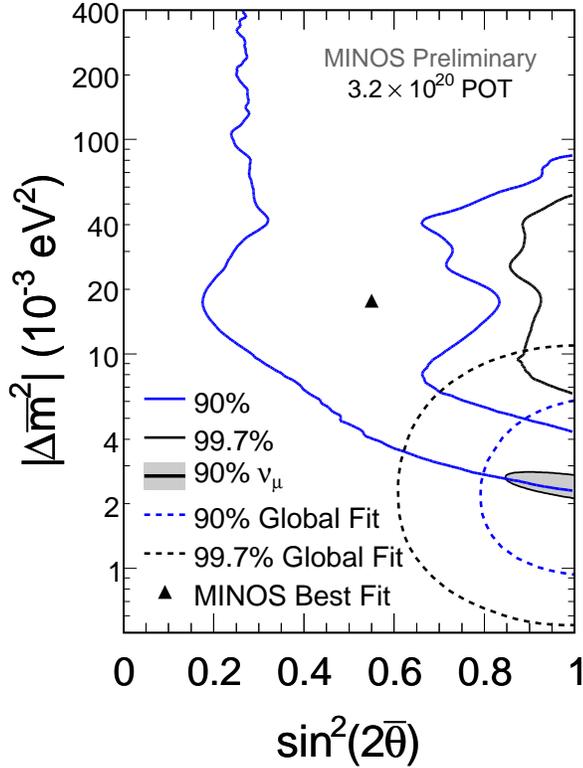}
\caption{Allowed oscillation-parameter regions from fit to the ${\bar \nu}_{\mu}$ data (solid contours) together with limits from a global fit to prior data (dashed) and from the MINOS $\nu_{\mu}$ analysis (shaded).} 
\label{fig:contour}
\end{figure}

\begin{figure}[h]
\centering
\includegraphics[width=80mm]{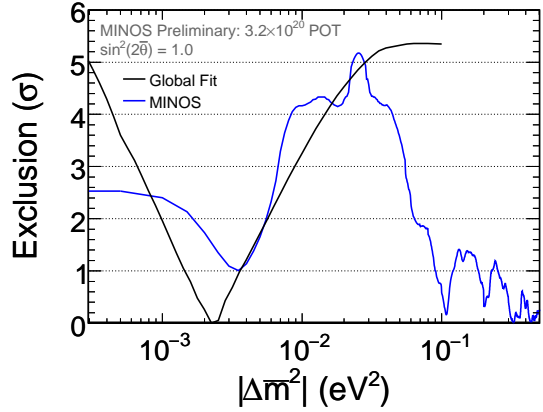}
\caption{Confidence level of one-parameter fit to the MINOS ${\bar \nu}_{\mu}$ data (blue) compared to that from the global fit (black).} 
\label{fig:onepar_fit}
\end{figure}

Fig.~\ref{fig:contour} shows the resulting contours on the muon antineutrino oscillation parameters, $\sin^2(2{\bar \theta})$ and $\Delta {\bar m}^2$, together with limits from a recent global fit to earlier data \cite{Global_fit} as well as MINOS limits on the corresponding neutrino oscillation parameters \cite{MINOS_CC}. Although the best fit point is at high $\Delta {\bar m}^2$ because of the $1.9\sigma$ overall deficit, the data are consistent with CPT-invariant oscillation parameters at $90$\% C.L. and exclude the no-oscillation scenario at $99$\% C.L. We also perform a 1-parameter fit at maximal mixing ($\sin^2(2{\bar \theta}) = 1$), which helps to rule out previously allowed regions of $\Delta {\bar m}^2$ (Fig.~\ref{fig:onepar_fit}). At maximal mixing, $\Delta {\bar m}^2 < 2.0 \times 10^{-3}$ and $5.1 \times 10^{-3} < \Delta {\bar m}^2 < 81 \times 10^{-3}$ are excluded at $90$\% conficence.

There is no evidence for excess of events at low energy that would indicate $\nu_\mu \to {\bar \nu}_{\mu}$ transition, and a fit to the data under this assumption gives an upper limit $\alpha < 2.6$\% at $90$\% C.L.

\section{Future prospects}

MINOS has already collected about $7.2 \times 10^{20}$ PoT data in the $\nu_{\mu}$-dominated beam configuration. The analysis of this data is under way and updated results are expected in the near future.

\begin{figure}[h]
\centering
\includegraphics[width=80mm]{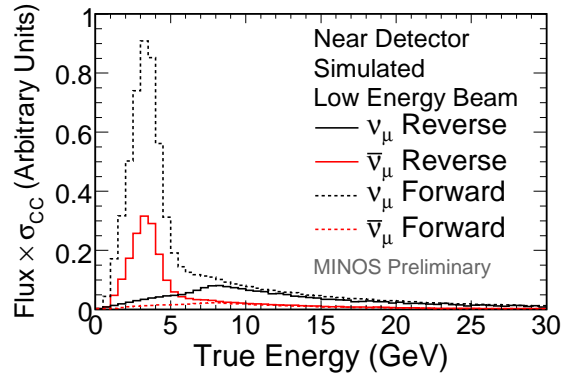}
\caption{The ${\bar \nu}_{\mu}$-CC (red) and $\nu_{\mu}$-CC (black) Near Detector spectra in antineutrino (solid) and neutrino (dashed) dominated beam.} 
\label{fig:RHC_spectra}
\end{figure}

More importantly for this analysis, the current of the NuMI focusing horns will be reversed in the Fall of 2009 in order to produce a dedicated antineutrino beam\footnote{The polarity of the magnetic horns was reversed and antineutrino running commenced on September 29, 2009.}. Fig.~\ref{fig:RHC_spectra} shows the expected ${\bar \nu}_{\mu}$ and $\nu_{\mu}$-CC spectra in the antineutrino mode relative the current neutrino mode. The ${\bar \nu}_{\mu}$-CC event rate per PoT in the antineutrino mode is about a factor of three lower than the $\nu_{\mu}$-CC event rate in the neutrino mode due to lower production rate of negative mesons in the target and about a factor of two lower ${\bar \nu}_{\mu}$-CC cross section in the detector. Despite this, the new configuration will significantly increase the ${\bar \nu}_{\mu}$-CC event rate, in particular in the low energy region which is more sensitive to the oscillation, compared to the current beam configuration.

\begin{figure}[h]
\centering
\includegraphics[width=80mm]{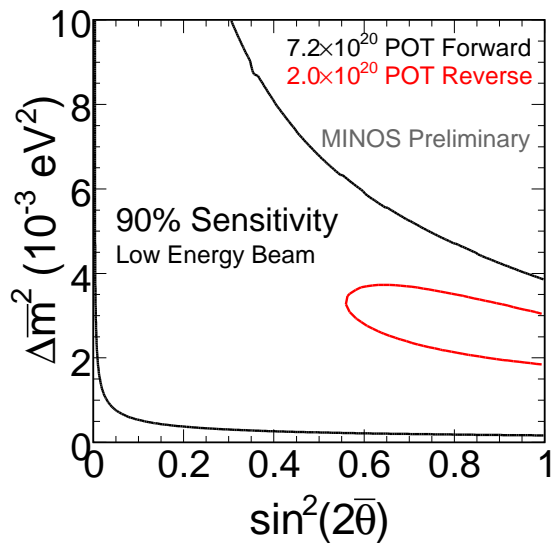}
\caption{Sensitivity to ${\bar \nu}_{\mu}$ oscillation parameters with antineutrino (red) and neutrino (black) dominated beam.}  
\label{fig:RHC_contour}
\end{figure}

Although the run plan is not yet finalized, we expect to collect about $2 \times 10^{20}$ PoT data in the antineutrino configuration in one year of running before the Summer shutdown in 2010\footnote{At the time of writing, the plan is to run in antineutrino configuration until March 1, 2010.}. Fig.~\ref{fig:RHC_contour} shows how much improvement can be achieved in sensitivity to the antineutrino oscillation parameters with $2 \times 10^{20}$ PoT data collected in antineutrino mode compared to the existing $7.2 \times 10^{20}$ PoT data in neutrino mode. This will allow us to measure the oscillation parameters, in particular $\Delta {\bar m}^2$, for muon antineutrinos with high precision and establish antineutrino oscillation at $5\sigma$ confidence level.

\begin{acknowledgments}
This work was supported by the US DOE; the UK STFC; the US NSF; the State and University of Minnesota; the University of Athens, Greece; and Brazil's FAPESP and CNPq. We are grateful to the Minnesota Department of Natural Resources, the crew of the Soudan Underground Laboratory, and the staff of Fermilab for their contribution to this effort.
\end{acknowledgments}

\bigskip 

\end{document}